\documentclass[prb,twocolumn,showpacs,preprintnumbers,amsmath,amssymb]{revtex4}

\usepackage{graphicx}
\usepackage{dcolumn}
\usepackage{bm}
\usepackage{revsymb}
\usepackage{txfonts}

\begin{document}

\preprint{cond-mat/0311069,  Last modified on 24 Janualy 2004.}

\title{STM observation of electronic wave interference effect in finite-sized graphite\\with dislocation-network structures}

\author{Yousuke Kobayashi}
 \email{ykobaya@chem.titech.ac.jp}
\author{Kazuyuki Takai}
\author{Ken-ichi Fukui}
\author{Toshiaki Enoki}
 \affiliation{Department of Chemistry, Tokyo Institute of Technology, 2-12-1, Ookayama, Meguro-ku, Tokyo 152-8551, Japan}

\author{Kikuo Harigaya}
 \affiliation{Nanotechnology Research Institute, AIST, Tsukuba 305-8568, Japan\\Synthetic Nano-Function Materials Project, AIST, Tsukuba 305-8568, Japan}

\author{Yutaka Kaburagi}
\author{Yoshihiro Hishiyama}
 \affiliation{Department of Energy Science and Engineering, Musashi Institute of Technology, 1-28-1, Tamazutsumi, Setagaya-ku, Tokyo 158-8557, Japan}


\begin{abstract}
 Superperiodic patterns near a step edge were observed by Scanning Tunneling Microscopy on several-layer-thick graphite sheets on a highly oriented 
pyrolitic graphite substrate, where a dislocation network is generated at the interface between the graphite overlayer and 
the substrate. Triangular- and rhombic-shaped periodic patterns whose periodicities are around 100 nm were observed on the 
upper terrace near the step edge. In contrast, only outlines of the patterns similar to those on the upper terrace were 
observed on the lower terrace. On the upper terrace, their geometrical patterns gradually disappeared and became similar 
to those on the lower terrace without any changes of their periodicity in increasing a bias voltage. By assuming a 
periodic scattering potential at the interface due to dislocations, the varying corrugation amplitudes of the patterns can 
be understood as changes in the local density of states as a result of the beat of perturbed and unperturbed waves, i.e., the interference in an 
overlayer. The observed changes in the image depending on an overlayer height and a bias voltage can be explained by the 
electronic wave interference in the ultra thin overlayer distorted under the influence of dislocation-network structures.
\end{abstract}

\maketitle

\section{\label{sec:level1}Introduction}

 Scanning tunneling microscopy (STM) observations of superperiodic patterns on metal surfaces have been reported in several 
finite-sized systems. They are ascribed to interference patterns of free electron waves scattered by adatoms and step edges, 
for example, an Ag(111) surface near a step edge,\cite{ref1,ref2,ref3,ref4} a Cu(111) surface surrounded by 76 Fe adatoms,\cite{ref4,ref5,ref6} 
and so on. These reports have clarified that scattered and interfered waves on the surface can be observed as periodic patterns which are related to 
the Fermi surface of bulk and surface states of metals where free electrons can move around. Recently, a superperiodic 
pattern has been also reported in semiconductor surfaces such as InAs/GaAs(111)\textit{A}.\cite{ref7,ref8} In this case, the pattern is also an 
interference pattern, which is generated by electron waves scattered at step edges on semiconductor surfaces because a 
two-dimensional (2D) electron gas is generated due to the band bending by the surface reconstruction. This phenomenon is 
interesting and characteristic of the surface electronic structure of isotropic semiconductors; that is, generated electrons 
whose characters resemble free electrons in metals can move in a surface thin layer in spite of three dimensionality in the 
electronic structures of semiconductors. As for 2D electronic systems, the present authors have observed the electronic wave 
interference effect on nanographene sheet inclined with respect to a highly oriented pyrolytic graphite (HOPG) substrate by 
STM.\cite{ref9} Here, a nanographene sheet interacts very weakly with the HOPG substrate, where electrons are confined in the 2D sheet 
and the in-plane potential changes gradually.\\
\hspace*{10pt}Meanwhile, superperiodic patterns on an HOPG substrate observed by STM have been also reported in many papers.\cite{ref10,ref11,ref12,ref13,ref14,ref15,ref16,ref17,ref18} Those 
patterns are not generated by electronic wave interference effects, but are caused by multiple-tip effects, rotational 
stacking faults, and dislocation-network structures. A pattern caused by multiple-tip effects originates from superimposing 
two different information of graphite lattice in one domain imaged by a tip apex and that in another domain, with a relative 
rotation, imaged by a metal contamination that is attached to the tip. A moir\'{e} pattern and a pattern caused by the 
dislocation-network structures result from the spatially varied local density of states (LDOS), which are related to the 
stacking faults by the relative rotation between two adjacent graphene layers and by the lattice distortion at the interface, 
respectively. Graphite with a stacking fault can be represented as $abcab\dots$ where represented as $ababa\dots$ for an ordinary stacking of graphite 
and $c$ for a faulted layer. The periodicity of a moir\'{e} pattern and a pattern caused by the dislocations can be explained by an angle of the relative rotation\cite{ref11,ref12,ref13,ref14} and by the 
periodic domain of stacking faults generated in a slip plane, respectively. Among the reports, a change in the bias voltage interestingly induces a change in the periodicity of the 
superperiodic patterns that come from the dislocation network, similar to that is observed by transmission electron microscopy (TEM).\cite{ref15}\\
\hspace*{10pt}LDOS calculation of faulted stacking parts cannot reproduce the corrugation amplitude of superperiodic structures that have been reported so far.\cite{ref13,ref19,ref20} About a moir\'{e} pattern 
and a dislocation-induced pattern, one can find that the interface between an overlayer and a substrate is taken as a scattering layer and that the overlayer is regarded as a 
finite-sized region in the normal direction to the surface.\cite{ref21} Electron waves normal to the surface can be scattered by the surface and the interface, resulting in the generation 
of standing waves. This is the electron confinement effect in the 1D direction normal to the surface, which is very important for the superperiodic LDOS at the surface in 
terms of the corrugation amplitude in a STM image. The corrugation amplitude of a superperiodic pattern is expected to depend on a bias voltage of STM and the overlayer 
thickness as the character of waves in the overlayer. In this paper, we report on the observation of different superperiodic patterns that originate from the dislocation-network 
structures, on both terraces near a step edge and present their bias voltage dependence of the corrugation amplitudes with no change in the periodicity. The patterns are 
explained as the spatially varied LDOS affected by the interference in the overlayer.

\section{\label{sec:level2}Experimental}

 All images in the present paper were observed by using a commercial STM system (Digital Instruments, Nanoscope E) under an 
ambient condition at room temperature with the constant-current mode at 0.7 nA using a mechanically cut Pt-Ir tip. Sample 
bias voltages for these observations were varied from near the Fermi level to higher voltages, typically at 0.02, 0.05, 0.1, 
0.2, 0.3, 0.4, 0.5, and 0.6 V, except for that in Fig.2(b). The sample was fabricated by the heat-treatment of an HOPG 
substrate at 1600 \symbol{"17}C in Ar flow after cleaving it by an adhesive tape for obtaining a fresh surface. It is possible that 
dislocations were generated at several layers beneath the surface during the heat-treatment process.

\section{\label{sec:level3}Results}

\begin{figure}
\includegraphics[width=7.2cm, clip]{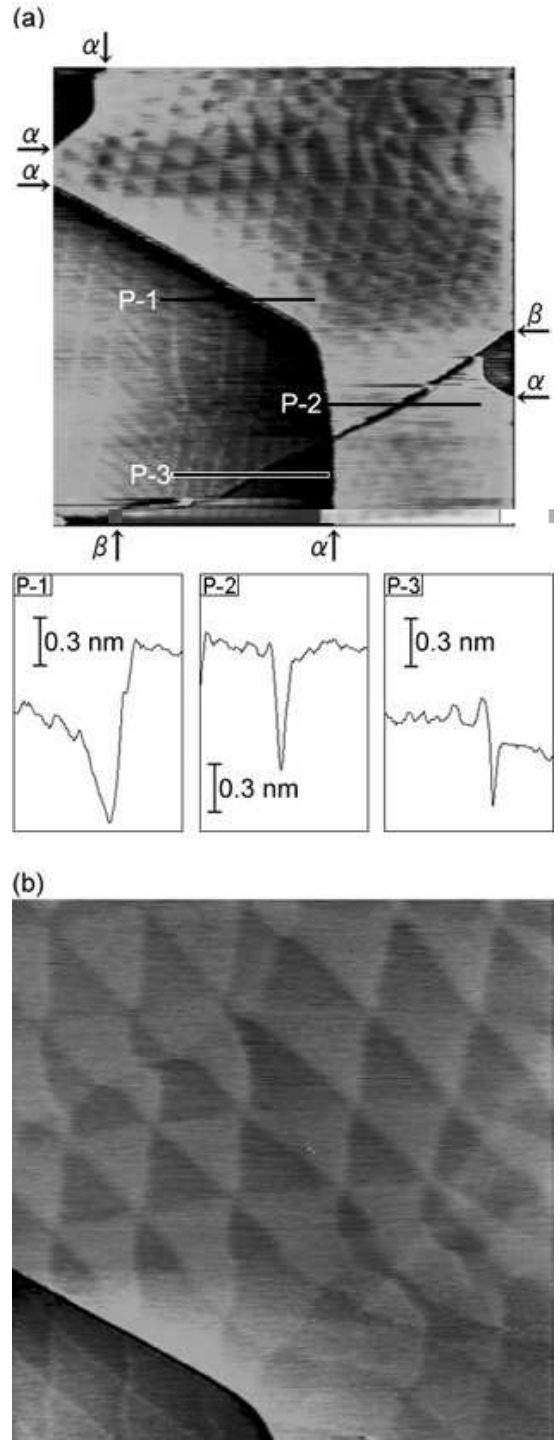}
\caption{\label{fig:inversion1} (a) STM image (1.5$\times $1.5 $\mu $m$^2$) of superperiodic patterns observed at $V_s${}=0.2 V. There are 
triangular-, rhombic- and net-shaped patterns whose periodicity is around 100 nm. There are also complicated patterns in 
some parts, where two types of patterns are superimposed. Arrows $\alpha $ and $\beta $ denote the positions of a step edge of graphite 
and a domain boundary, respectively. Cross-sectional profiles are taken along the horizontal lines, P-1 to P-3, whose 
lengths are 500 nm. (b) Magnified STM image (500$\times $500 nm$^2$) of the center region of (a), showing two types of patterns. The 
heights of the lower and upper terraces correspond to two and three graphene layers from the substrate, respectively.}
\end{figure}
 Superperiodic patterns were observed on the surface by STM, which extended over several $\mu$m$^2$, and a part of the area is shown 
in an 1.5$\times$1.5- $\mu$m$^2$ image in Fig.1(a), where a bias voltage was 0.2 V. In this image, there are triangular-, rhombic-, and 
net-shaped patterns whose periodicities are around 100 nm but gradually changed depending on the position. There are also 
complicated patterns that seem to be the superimposed of those patterns. The lines pointed by arrows $\alpha$ denote step edges of 
graphite, while the line pointed by arrows $\beta$ denotes a domain boundary where the difference in the heights between the two 
regions faced at the boundary is much less than the interlayer distance of graphite (0.335 nm in the bulk). The presence of the 
step edge and the domain boundary is confirmed by the cross-sectional profiles shown in Fig.1(a). The lower terrace in 
Fig.1(a) extends to the left direction by about 3.5 $\mu $m and is terminated by a boundary between the graphene overlayer and the HOPG 
substrate. Patterns of those shapes change into y-shaped and linear patterns near the graphene overlayer edge and end at the 
edge (the region showing the y-shaped and linear patterns is not shown). The triangular- and net-shaped patterns similar to 
those in Fig.1(a) have been previously observed in STM,\cite{ref12,ref13} TEM,\cite{ref22,ref23,ref24} and other investigations.\cite{ref25,ref26,ref27} In those reports, 
diffraction patterns in TEM images and superperiodic patterns in STM images were attributed to the modified LDOS caused by 
rhombohedral stacking faults due to partial dislocations. The partial dislocations are defined by the Burgers vector that 
converts an $ab$-stacked layer in ordinary graphite to an $ac$-stacked layer with respect to a glide plane. The conversion of 
stacking occurs abruptly accompanied with a lattice distortion where a sharp-edged periodic pattern is generated. Therefore, 
the patterns in Fig.1(a) are considered to come from dislocations at the interface between the graphite overlayer and the 
HOPG substrate from the shapes and the average periodicity.\\
\hspace*{10pt}A magnified image of Fig.1(a) near a step edge is shown in Fig.1(b). The height of the lower terrace at the bottom left in 
Fig.1(b) from the HOPG substrate is 0.67$\pm$0.02 nm in average (for $V_s$=0.2 V) from the cross-sectional profile analyses 
of the observed image at the boundary between the graphene overlayer and the substrate (not shown). The value corresponds to 
a thickness of two graphene layers from the substrate. The upper terrace at the center part in Fig.1(b) has a height of 
three graphene layers from the substrate, as estimated from the cross-sectional profile of the step edge, whose height 
difference is 0.39-0.41 nm (for $V_s${}=0.2 and 0.5 V) including the corrugation amplitude of superperiodic patterns [P-1 in 
Fig.1(a)]. The image at a low bias voltage of 0.02 V near the Fermi energy is shown in Fig.2(a). Though this image was obtained 
at almost the same place as that shown in Fig.1(b), there are a few differences in the contrast and the shape of the patterns. 
In Fig.2(a), three regions are indicated; regions A, B, and C contain a triangular-shaped pattern, a rhombic-shaped pattern, 
and a net-shaped pattern, respectively. In intermediate regions A-B and B-C, 
\begin{figure}
\includegraphics[width=7.8cm, clip]{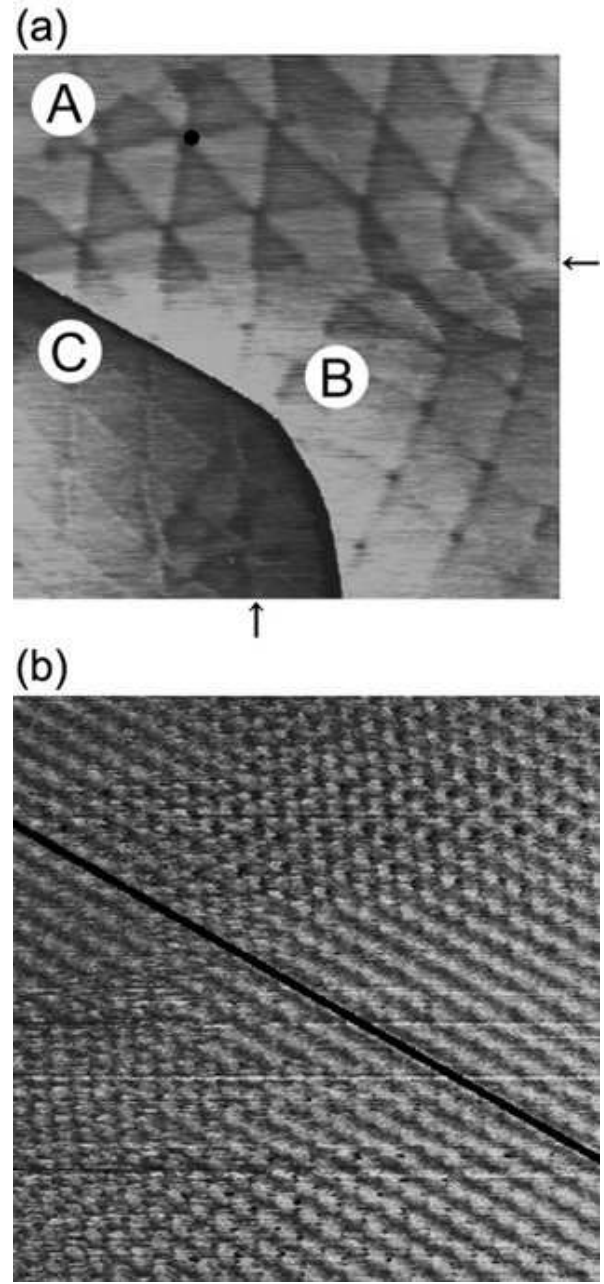}
\caption{\label{fig:inversion2} (a) STM image (500$\times $500 nm$^2$) of the superperiodic patterns at a low sample bias voltage of 0.02 V, 
which is expected to reflect the density of states close to the Fermi level. The imaged area is shifted from that of Fig.1 
to the bottom by about 200 nm. A, B, and C denote regions of triangular-, rhombic- and net-shaped patterns, respectively. 
Arrows indicate complicated patterns, where two patterns are superimposed. Lines that divide the geometric patterns into 
individual units cross at contracted nodes in regions A and B, and at extended nodes in region C. The apparently depressed 
contrast neighbors to the lines are artificial effect to make the image clearer in region C. (b) Atomically resolved STM 
image (6.0$\times $6.0 nm$^2$) of one individual triangular pattern near a contracted node on the upper terrace in (a), which is marked 
by a black dot, at $V_s$=0.002 V and $I$=1.7 nA. A straight line placed on triangular lattice sites at the bottom right part is 
extended to the valley sites of the triangular lattice at the top left part through a distorted lattice part.}
\end{figure}
there are complicated contrasts that are 
\begin{figure*}
\includegraphics[width=15cm, clip]{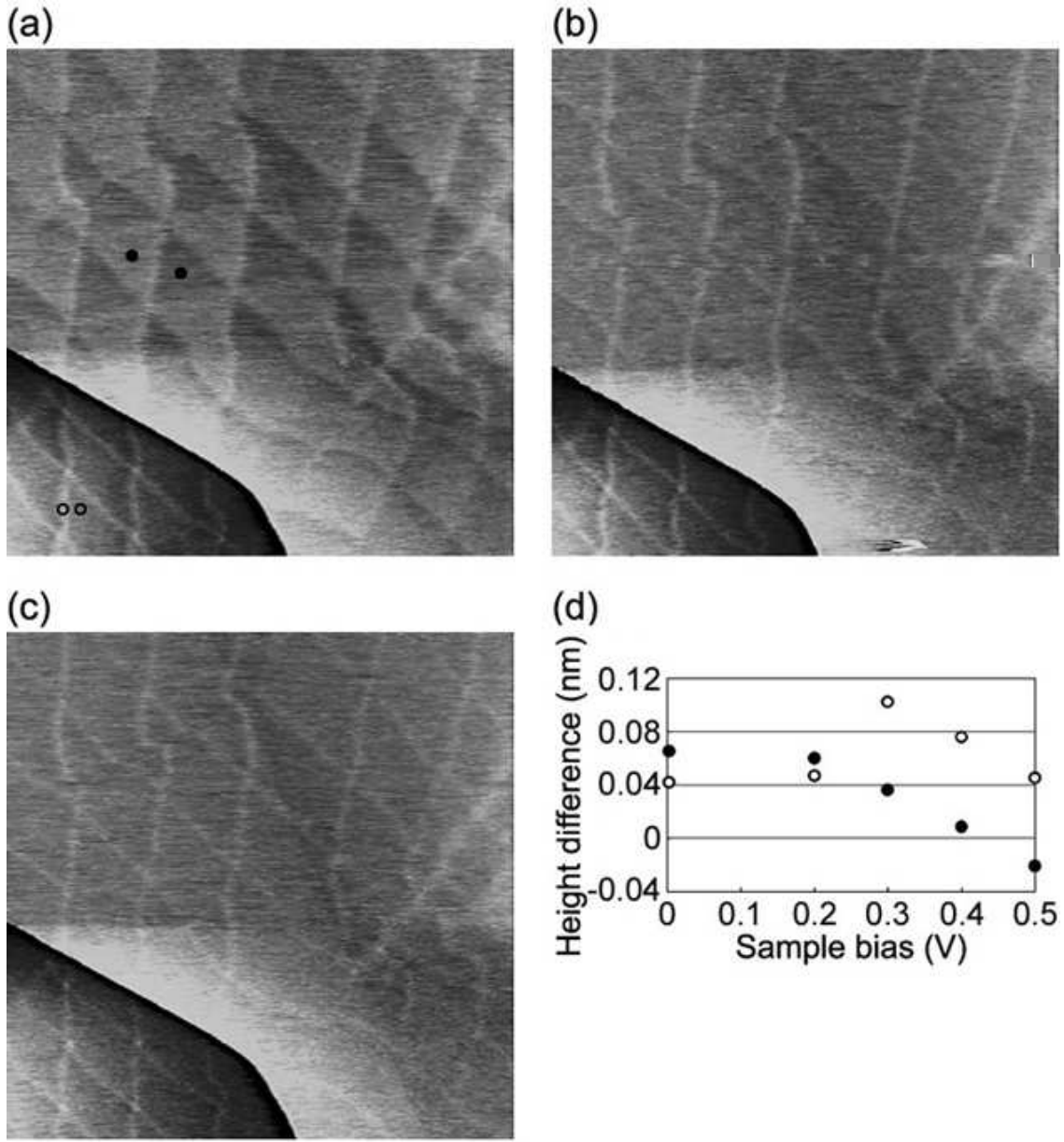}
\caption{\label{fig:inversion3} STM images (500$\times $500 nm$^2$) of superperiodic patterns at higher sample bias voltages; (a) $V_s$=0.3 V, 
(b) $V_s$=0.4 V, and (c) $V_s$=0.5 V. By increasing the bias voltage, the corrugation amplitude of superperiodic patterns on 
the upper terrace decreased gradually [(a) and (b)] and changed into a net pattern (c). In contrast, no significant change was 
observed for the pattern on the lower terrace. The net pattern appearing on the upper terrace of (c) is similar to that on 
the lower terrace. Height differences between two points depicted in (a) are shown in (d) for clarifying the bias-dependent 
contrast. Solid and blank circles are the height differences of the upper and lower terrace, respectively. (Circles at the 
sample bias of around 0 V are the height differences at $V_s$=0.02 V.)}
\end{figure*}
superimposed of patterns in two regions. In regions A and B, the apparent height of lines which divide the patterns into 
individual geometric units was lower than the center of the unit by about 0.1 nm. Crossed points of lines were further 
depressed from the lines by about 0.1 nm, resulting in the ``contracted nodes'' in the image. In region C, however, lines are 
imaged higher than the center of the unit by about 0.05 nm and crossed points of the lines are the highest (about 0.005 nm 
higher than the lines), giving the ``extended nodes'' in the image. Except for the slight contrast, the patterns in regions B 
and C appear to have contrast inverted from each other. In Fig.1(b), however, the part corresponding to region B is not the 
same pattern as region B in Fig.2(a), suggesting that the corrugation amplitudes and the shapes of patterns depend on a bias 
voltage. The triangular shape in region A is almost the same as can be seen in the comparison of Fig.1(b) and Fig.2(a). 
However, the contracted nodes on the upper terrace at a bias voltage of 0.02 V [Fig.2(a)] changed to the extended nodes at 
0.2 V [Fig.1(b)]. Figure 2(b) is a magnified image near a contracted node of the upper terrace in Fig.2(a), which is marked 
by a black dot, taken at $V_s$=0.002 V, $I$=1.7 nA. A straight line drawn on triangular lattice points at the bottom right 
part is extended to the valley sites of the triangular lattice at the top left part, indicating the presence of a distortion 
at the center part of the image. This atomically resolved image supports that the observed patterns come from the 
dislocation-network structure. As for patterns at higher bias voltages, Figs.3(a)-(c) display images of almost the same 
places as Figs.1(b) and 2(a) at bias voltages of 0.3 V, 0.4 V, and 0.5 V, respectively. The patterns in Fig.3(a) seem to resemble 
those in Fig.1(b), except for the pattern change from rhombic-shaped to triangular-shaped in region B. In regions A and C, 
the corrugation amplitudes of patterns on the upper terrace in Fig.3(a) are smaller by about 0.03 nm than those in Fig.1(b), 
whereas those on the lower terrace in Fig.3(a) are larger by about 0.06 nm than those in Fig.1(b), as shown in Fig.3(d). For 
clarity, the height differences between two points on each terrace dependent on the bias voltage is shown in Fig.3(d). Just 
by increasing a bias voltage, patterns on the upper terrace in Fig.3(a) are changed into a net-shaped pattern on the lower 
terrace in Fig.3(c). They are similar to the pattern on the lower terrace in Figs.3(a)-(c), however, slightly inversed 
contrast is observed as shown in Fig.3(a) similar to that in region C in Fig.2(a). Changes of corrugation amplitudes, with a maximum at $V_s$=0.3 V are observed in 
the patterns on the lower terrace as shown in Fig.3(d). A similar image to Fig.3(c) was also observed at $V_s$=0.6 V (not shown).

\section{\label{sec:level4}Theoretical model and discussion}

 According to previous reports, the periodicity of superperiodic patterns changed dependent on a bias voltage or just by 
scanning the tip, which was attributed to dislocation motion.\cite{ref15, ref16} However, the periodicity of the observed 
patterns in the present study did not change in the range of voltages used for imaging (0.02 - 0.6 V). Therefore, the 
observed phenomenon is different from that in Ref.15 and 16, which show a dislocation motion induced by the applied bias 
voltage. A dislocation motion is not generated in the observed phenomenon in the present paper. Instead, only the 
superperiodic corrugation amplitudes of the observed patterns varied, without any change in the periodicity, depending on 
an overlayer height from the substrate and a bias voltage of STM, as shown in Fig.3(d). This is the first observation of 
the bias-dependent contrast and pattern shapes of superperiodic patterns on graphite without any change of the periodicity.//
\begin{figure}
\includegraphics[width=5cm, clip]{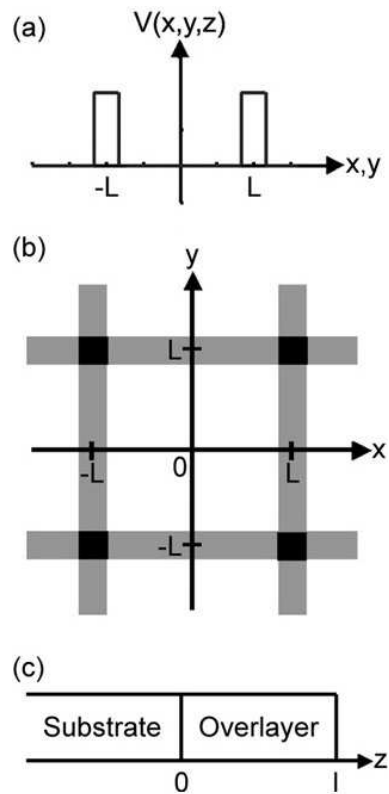}
\caption{\label{fig:inversion4} Model of the potential at the interface between the graphite overlayer and the substrate. (a) 
The cross-sectional profile of square potential along the $x$ or $y$ axis. The periodicity is 2$L$ and the potential is $L$/3 
in width and 2$v_0\delta (z)$ in height. (b) The projection of the square-patterned potential on the $xy$ plane. Gray lines represent 
potential lines and black squares represent potential nodes, whose potential height is the sum of the 1D potentials in $x$  
and $y$ axes. A height of the potential nodes is twice as large as that of the potential lines. (c) The position of the 
surface and the interface along the $z$ axis. The surface and the interface are located at $l$ and 0 in the $z$ axis, respectively.}
\end{figure}
\hspace*{10pt}First, we discuss the origin of the bias-dependent corrugation amplitudes. As shown in Figs.1-3, it appears that the 
patterns on the upper and lower terraces are connected continuously at the step edge, independent of the bias voltage. 
This suggests that the patterns observed on both terraces come from the same origin. It should be noted that the 
experimental results cannot be explained simply by calculating the DOS of faulted stacking, because the observed patterns 
at different terraces have contrast inverted from each other as shown in Figs.1-3(b). It seems natural to assume that an 
array of faulted stacking is not changed abruptly across the step edge if the dislocation network is continuous at the 
interface. We cannot also explain the property, on the basis of the faulted stacking, that the superperiodic corrugation 
amplitudes on the lower terrace become larger although the gap between the tip and the sample becomes larger in increasing 
a bias voltage from 0.02 to 0.3 V as shown in Fig.3(d). In other words, the observed behavior is considered to be due to 
the LDOS at the surface, taking into account the fact that observed corrugation amplitudes on the lower terrace become larger. Then, the LDOS 
should explain the gradual decreases of the corrugation amplitudes and the variations of patterns on the upper terrace in 
increasing the bias voltage without changing the periodicity of the patterns, and that should also explain the increase of 
the corrugation amplitudes on the lower terrace in increasing a bias voltage near the Fermi energy.\\
\hspace*{10pt}Here, we discuss the interference effect for explaining the bias-voltage dependence of superperiodic patterns on the basis 
of a theoretical treatment reported in Ref.21. Considering the scattering potential at the interface, one can find that the 
LDOS at the surface is related to the interference effect of electrons that are scattered at the surface and the interface 
between the overlayer and the substrate. The LDOS at the surface can be given as $\sin^2(kz)$ using coordinate $z$ and a 
wave number $k$ along the axis normal to the surface (the $z$ axis) in case that the lateral wave number of a 
superperiodic pattern nearly equals to 0 by comparison with the wave number originating from the lattice. If one treats a 
scattering potential at the interface by perturbation, a beat can be generated by the interference between the perturbed 
and the unperturbed waves. In this case, the LDOS at the surface is proportional to $\sin(kz)\cos(k'z)$, where $k'$ and $k$ 
are a perturbed wave number and an unperturbed wave number, respectively. Next we will show the detailed derivation.\\
\hspace*{10pt}In the present discussion, a square-patterned potential with a periodicity of 2$L$ at the interface is employed as shown in 
Fig.4, for a calculation of the probability density of the wave function confined in the plane for generating an abrupt 
potential change associated with the dislocation-network structures. We place a square potential with $L$/3 in width and 
2$v_0$$\delta(z)$ in height, where $L$ is the half of the periodicity of the square potential and $v_0$ is the strength of the 
scattering potential, at the line dividing the patterns into geometrical units as a simple model to reproduce the patterns 
in regions B and C, as shown in Fig.4(a) and (b). Though the square-shaped pattern in the present model is different from the 
experimental result (the rhombic-shaped pattern in regions B and C, the triangular-shaped pattern in region A, and the complicated pattern in their 
intermediate regions), it can make a theoretical treatment easier with any loss of validity. (Note that the problem of 
the square-patterned potential in rectangular coordinate can be reduced to the 1D problems along the $x$ and $y$ axes.) 
The reproduction of a pattern in region A is beyond the present model since the shape of a pattern in region A is 
complicated to solve in the similar manner that is applied for regions B and C. If we locate the surface and the interface 
positions at $l$ and 0, respectively, in the $z$ axis as shown in Fig.4(c) and introduce the delta function $\delta (z)$ at the 
interface, this potential can be expressed using the Fourier analysis
\begin{multline}
V(x,y,z)=(\hbar^2/m_\bot)v_0\sum_{n} a_n\delta(z)
 \\ \times (\mathrm{e}^{\mathrm{i}q_{xn}\cdot x}+\mathrm{e}^{-\mathrm{i}q_{xn}\cdot x}+\mathrm{e}^{\mathrm{i}q_{yn}\cdot y}+\mathrm{e}^{-\mathrm{i}q_{yn}\cdot y}),
\end{multline}
where $\hbar$ is the Planck constant over 2$\pi $, $m_\bot$ is the effective mass along the $z$ axis, an is the $n$th component which equals 
to $\{2/(n\pi)\}\{\sin(n\pi)-\sin(5n\pi/6)\}$ for the square potential, and $q_{xn}$ and $q_{yn}$, which take discrete values $(n\pi/L)$
($n$=$1,2,\dots$), are the $n$th wave vectors in the $x$ and the $y$ axes, respectively. Assuming the linear combination of in-plane 
plane waves and wave function $A_{q_x ,q_y}(z)$ for the $z$ component, the wave function is represented to be
\begin{equation}
\Psi (x,y,z)=\sum_{q_x ,q_y} A_{q_x ,q_y} (z)\mathrm{e}^{\mathrm{i}(q_x \cdot x+q_y \cdot y)}.
\end{equation}
Based on this wave function and the connecting condition derived from the Schr\"{o}dinger equation with the square-patterned 
potential, the perturbed wave function is given as
\begin{multline}
\Psi ^\pm (x,y,z)=v_0 \sum_{n} \{a_n / (\mathrm{i}k')\}(1-\mathrm{e}^{2\mathrm{i}kl})\\ \times (\mathrm{e}^{\mathrm{i}k'z}-\mathrm{e}^{-\mathrm{i}k'z+2ik'l})\mathrm{e}^{\pm (q_{xn}\cdot x+q_{yn}\cdot y)},
\end{multline}
where the term of $v_0 ^2$ is neglected because of its small contribution. For the unperturbed wave, 
we take a plane wave in the direction of the $z$ axis,
\begin{equation}
\Psi _0 =\mathrm{e}^{\mathrm{i}kz},
\end{equation}
with
\begin{equation}
k'^2 =k^2 -(m_\parallel /m_\perp )\mid q_n \mid ^2,
\end{equation}
where $m_\parallel$ is the effective mass in the $xy$ plane. In the overlayer shown in Fig.4(c), 
the unperturbed waves can coexist with the perturbed waves. Since the total wave function  
$\Psi _{\mathrm{total}}(x,y,z)$  is the sum of $\Psi _+$, $\Psi _-$, and $\Psi _0$, the probability density, $\mid \Psi _{\textrm{total}}(x,y,z)\mid ^2$ 
is represented as follows:
\begin{multline}
\mid \Psi _\textrm{total}(x,y,z)\mid ^2 =4\sin^2(kz)\\
-32\sum_{n} (a_nv_0/k')\sin(kl)\cos(k'l)\\
\times \sin(kz)\sin(k'z)\\
\times \{\cos(q_{xn}\cdot x)+\cos(q_{yn}\cdot y)\}.
\end{multline}
Here, the $xy$ plane is shifted from ($z-l$) to $z$ for simplicity and the very small term of $v_0^2$ 
is also neglected. Using the Tamm states near the surface ($z\sim 0$), $k\cot(kz)$, and $k'\cot(k'z)$ can be 
regarded as the constant numbers, provided that $kz\sim 0$ and $k'z\sim 0$. Therefore, the probability density 
around the surface, $\mid \Psi_\textrm{total}(x,y,0)\mid ^2$, is roughly expressed with an overlayer height from the 
substrate, $l$, the perturbed wave number, $k'$, and the unperturbed wave number, $k$:
\begin{multline}
\mid \Psi _\textrm{total}(x,y,0)\mid ^2 =-c\sum_{n} a_nk\sin(kl)\cos(k'l)\\
\times \{\cos(q_{xn}\cdot x)+\cos(q_{yn}\cdot y)\}+\textrm{const.},
\end{multline}
where $c$ and the second term are positive constants and the second term is larger absolute value than that of the first 
term. In this equation, the spatially varied probability density, which gives a superperiodic 
pattern, corresponds to the term, $\{\cos(q_{xn}\cdot x)+\cos(q_{yn}\cdot y)\}$. By attributing the unperturbed wave to the wave function in 
bulk graphite, the energy dispersion can be given using parameter $m_\perp $, the interlayer distance $c$, and the interlayer resonance integral, 
$\gamma _1$(=0.39 eV) (Ref.28):
\begin{equation}
E=\hbar^2k^2/(2m_\perp )-2\gamma _1,
\end{equation}
with
\begin{equation}
m_\perp =\hbar ^2/(2c^2\gamma _1).
\end{equation}
The investigation of the spatially varied LDOS is important in order to look over the contrast 
image of STM from the corrugation amplitude of a superperiodic pattern that depends on a bias 
voltage. In this connection, the difference of the LDOS at the surface, 
\mbox{$\{\mid \Psi _\textrm{total}(0,0,0)\mid ^2-\mid \Psi _\textrm{total}(L,0,0)\mid ^2\}$}, where the former and latter terms represent the LDOS at 
the center and edge of an individual geometric pattern unit, respectively, can give a simple 
diagnosis in mapping a superperiodic pattern because a potential height is constant except for the 
edge part with a fine oscillation resulted from the Fourier analysis. In Fig.5, the difference of 
the LDOS in an arbitrary unit varies as a function of an overlayer height from the interface and a 
bias voltage. It obviously changes in the present range of an overlayer height from the interface 
and a bias voltage that were used for STM observation. The positive value of the difference means 
that the LDOS at the bias voltage is larger at the center position, $(x,y)$=$(0,0)$, surrounded by 
four potential lines than on the potential line, $(x,y)$=$(L,0)$, giving a square-shaped pattern. 
Conversely, the negative value suggests that the LDOS at the center position is smaller than 
that on the potential line, the superperiodic pattern being a net-shaped pattern. As the absolute value 
of the difference is proportional to the superperiodic corrugation amplitudes, the increase, 
decrease, and inversion of a corrugation amplitude of the superperiodic pattern can be generated 
depending on a bias voltage and an overlayer height from the substrate, as shown in Fig.5. For a 
comparison between the theoretical model and the experimental results, an overlayer height from the 
interface, $l$, is given in the unit of a single graphene layer thickness, which corresponds to the 
interlayer distance of bulk graphite (0.335 nm). Assuming that the interface is located at the 
intermediate plane between the overlayer and the substrate (an ideal interface is shifted by a 
half of monolayer from the substrate), $l$ is given as
\begin{figure}
\includegraphics[width=6cm, clip]{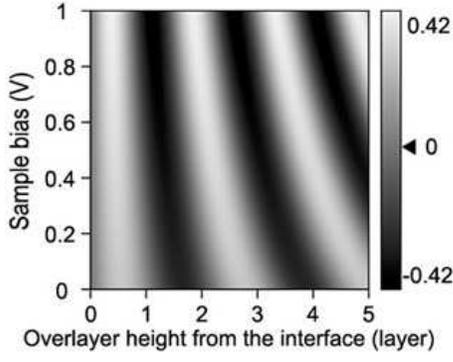}
\caption{\label{fig:inversion5} The difference of LDOS at two points, \mbox{\{$\mid \Psi _\textrm{total}(0,0,0)|^2-\mid \Psi _\textrm{total}(L,0,0)\mid ^2$\}}, 
as a function of an overlayer height from the interface and a bias voltage. The periodicity of a square potential, 2$L$, is 
70 nm. The difference of the LDOS is shown in scale bar and its unit is arbitrary. When the difference is a positive 
(negative) value, a calculated superperiodic pattern is a square-shaped (net-shaped) pattern. The unit number of the 
overlayer height (the number of layers) corresponds to the interlayer distance between adjacent graphene layers (0.335 nm).}
\end{figure}
\begin{equation}
l=s-0.5+\Delta ,
\end{equation}
where $s$ is an overlayer height from the substrate in the same unit as that of $l$, and $\Delta $ is 
a fitting parameter for a wavy structure of graphite.\cite{ref21}\\
\hspace*{10pt}Figure 6 shows the calculated LDOS in a $2L\times 2L$ square of the individual geometrical pattern unit at 
different overlayer heights and varied bias voltages. The extended and contracted nodes appear at 
the crossing points $(x,y)$=$(\pm L,\pm L)$,$(\pm L,\mp L)$. At $l$=1.5($s$=2.0, $\Delta$=0), the difference 
of the LDOS enhances with increasing a bias voltage from 0.02 to 0.3 V, where the calculated 
superperiodic patterns are net-shaped patterns with the extended nodes as clearly seen in 
Figs.6(a)-(d). 
\begin{figure}
\includegraphics[width=8cm, clip]{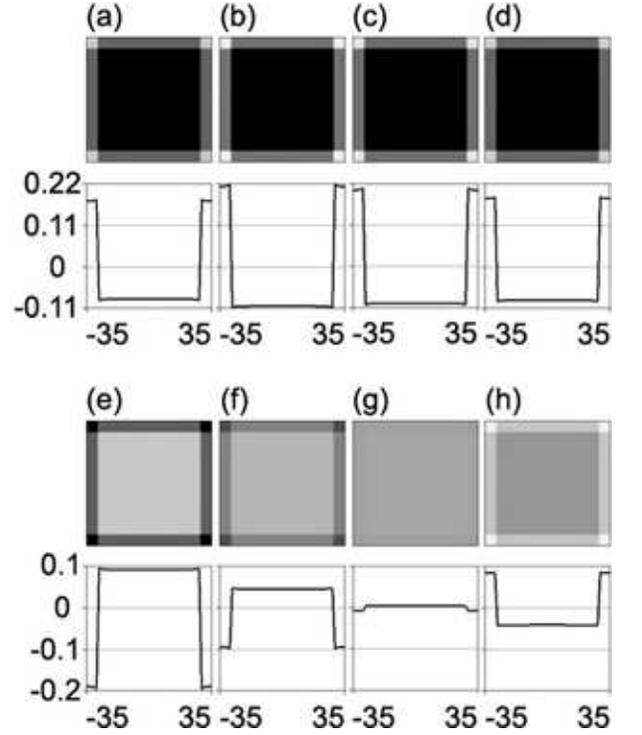}
\caption{\label{fig:inversion6} The calculated LDOS in a $2L\times 2L$ square (upper) and its cross-sectional profile 
passing through the center of the geometrical unit (lower) at different overlayer heights and bias voltages: (a) 
$l$=1.5 (layer), $V_s$=0.02 (V), (b) $l$=1.5, $V_s$=0.30, (c) $l$=1.5, $V_s$=0.40, (d) $l$=1.5, $V_s$=0.50, (e) $l$=2.7, $V_s$=0.02, (f) 
$l$=2.7, $V_s$=0.30, (g) $l$=2.7, $V_s$=0.40, and (h) $l$=2.7, $V_s$=0.50. (Top pictures) Lighter brightness indicates a higher 
LDOS value. (Bottom pictures) The $x$ or $y$ value in the lateral axis (unit: nm), the LDOS in the vertical axis (arbitrary units).}
\end{figure}
This result agrees with the experimental evidence that the inverted superperiodic 
patterns (net-shaped patterns) were observed on the lower terrace (two graphene layers high from 
the substrate) near the step edge, as shown in Figs.1-3. Indeed, the increase of the corrugation 
amplitude in Figs.2(a) and 3(a) can be understood on the basis of the increased difference of the LDOS 
because the corrugation amplitude is roughly proportional to the LDOS, as mentioned above. At 
$l$=2.7($s$=3.0, $\Delta$=0.2), the difference of the LDOS decreases in increasing a bias voltage from 
0.02 to 0.4 V, where the calculated superperiodic patterns become square-shaped patterns with 
contracted nodes as shown in Figs.6(e)-(g), although the pattern shape (square shaped) is 
different from that of experimental results (rhombic and triangular shaped). By further increase 
in the bias voltage (0.4 to 0.5 V), the difference of the LDOS has a negative value with a 
net-shaped pattern with extended nodes in Fig.6(h). This result supports the experimental evidence 
that the corrugation amplitude of the superperiodic pattern on the higher terrace (three graphene 
layers high from the substrate) decreases gradually and that the superperiodic pattern changes into an inverted pattern 
with increasing a bias voltage further, as shown in Fig.3. Eventually, the observed superperiodic 
patterns can be explained by the dislocation network at the interface and an interference in the 
overlayer dependent on its thickness and bias voltages.\\
\hspace*{10pt}However, this model cannot explain the relation between patterns in regions A and B, and the change 
in the corrugation amplitude of extending and contracting nodes. The potential, independent of the 
bias voltage in the present model, may be responsible for the discrepancy. A more appropriate way 
of faulted stacking or a slight relaxation is expected to improve the model, including a change of 
pattern shapes. The simple model used in this paper suggests that nodes can be alternated between 
the extended and the contracted due to the LDOS affected by the interference in the overlayer 
without changing the way of stacking. Other remaining problems are the discrepancy between the 
semimetallic electronic structure of graphite and the present theoretical model, and a phase shift 
of electron waves due to the Coulomb repulsion in the overlayer. These problems call for further 
investigations in the near future.

\section{\label{sec:level5}Conclusion}

 Superperiodic patterns that come form the dislocation-network structure have been observed by STM, where the shape and the 
corrugation amplitudes change dependent on a bias voltage and an overlayer height from the substrate without any variation 
of their periodicity. Near a step edge, the dislocation network that causes patterns on the upper and the lower terraces 
seems to be continuous. By assuming the same scattering potential at the interface between both terraces, a perturbed wave 
that generates a superperiodic pattern in the plane and an unperturbed wave can interfere in the overlayer, and a pattern at 
the surface can be affected by the beat of their waves. On the basis of the free electron model with the effective mass, 
the corrugation amplitudes of the patterns, which are related to the LDOS, are found to vary depending on a bias voltage 
and an overlayer height, and the changed corrugation amplitude of a superperiodic pattern can be understood as a change of 
the LDOS originating from the interference in the overlayer.
\begin{acknowledgements}
 The authors are grateful to Prof. Katsuyoshi Kobayashi (Ochanomizu University), for fruitful discussion. They also thank to Dr. A. Moore for his 
generous supply of HOPG sample. The present work was supported partly by the Grant-in-Aid for 'Research for the future' 
Program, Nano-Carbon, and 15105005 from Japan Society for the promotion of Science. One of the authors (K.H.) acknowledges 
the financial support from NEDO via Synthetic Nano Functional Materials Project, AIST, Japan.
\end{acknowledgements}
\bibliography{apssamp}
\end{document}